\documentclass[10pt,letterpaper]{article}
\usepackage{amssymb}

\usepackage{cogsci}
\cogscifinalcopy
\usepackage{pslatex}
\usepackage{apacite}
\usepackage{float}
\usepackage{algorithm}
\usepackage{algpseudocode}
\usepackage{longtable}% for long tables
\usepackage{amsmath} 
\usepackage{booktabs}
\usepackage[load-configurations=version-1]{siunitx} % newer version
\usepackage{natbib}

\newcommand{\figureref}[1]{Figure~\ref{#1}}
\algdef{SE}[INPUT]{Input}{EndInput}{\textbf{Input: }}{}
\algdef{SE}[OUTPUT]{Output}{EndOutput}{\textbf{Output: }}{}

\title{A Minimalistic Representation Model for Head Direction System}

\usepackage{graphicx}
\usepackage{multirow}

\author{
    \begin{tabular}{c c c}
        \bf Minglu Zhao$^{1}$  \quad\quad &
        \bf Dehong Xu$^{1}$ \quad\quad &
        \bf Deqian Kong$^{1}$      \\
        \normalfont minglu.zhao@ucla.edu \quad\quad & \normalfont xudehong1996@ucla.edu \quad\quad &\normalfont deqiankong@ucla.edu
    \end{tabular}
    \\\vspace{-8pt}\\
    \begin{tabular}{c  c}
        \bf Wen-Hao Zhang$^{2,3}$ \quad\quad &
        \bf Ying Nian Wu$^{1}$ \\
         \normalfont wenhao.zhang@utsouthwestern.edu\quad\quad &
        \normalfont ywu@stat.ucla.edu
    \end{tabular}
    \\\vspace{-8pt}\\
    \begin{tabular}{c }
        $^1$ Department of Statistics, UCLA 
    \end{tabular}\\
    \begin{tabular}{c}
        $^2$ Lyda Hill Department of Bioinformatics, UT Southwestern Medical Center 
    \end{tabular}\\    
    \begin{tabular}{c}
        $^3$ O’Donnell Brain Institute, UT Southwestern Medical Center
    \end{tabular}
}

\begin{document}

\maketitle

\begin{abstract}
We propose a minimalistic representational model for the head direction (HD) system, a crucial component of spatial navigation in mammals. Our model leverages the symmetry of the rotation group $U(1)$ and the inherent circular geometry of the head direction. We develop fully connected and convolutional versions of the model, both aiming to learn a high-dimensional representation of head direction that captures essential properties of HD cells. Our learning method results in representations that form a perfect 2D circle when projected onto their principal components, reflecting the underlying geometry of the problem. We also demonstrate that individual dimensions of our learned representation exhibit Gaussian-like tuning profiles akin to biological HD cells. Our model achieves accurate multi-step path integration, effectively updating its internal heading estimate over time. These results demonstrate that simple symmetry-based constraints can yield biologically realistic HD representations, offering new insights into the computational principles underlying spatial navigation in mammals.

\end{abstract}
% \begin{keywords}
% U(1) rotation symmetry group, Group representation, Recurrent neural network, Path integration
% \end{keywords}

\section{Introduction}

Spatial navigation is a fundamental cognitive function shared across many species, from insects to humans. A critical component of this navigational system is the perception of direction, which allows animals to maintain a consistent representation of their orientation in the environment. In mammals, this perception of direction is primarily mediated by the head direction (HD) system, a network of neurons that collectively encode the animal's current head orientation relative to its environment \citep{Taube1990}.

HD cells, discovered in the rat's dorsal presubiculum \citep{Rank1984,Taube1990a}, exhibit a remarkable property: they fire maximally when the animal's head faces a specific direction in the horizontal plane, regardless of location or ongoing behavior. Each cell has a preferred direction, with firing rates decreasing as the head turns away, typically following a Gaussian-like tuning curve \citep{Blair1997}. Distributed across interconnected brain regions \citep{Taube2007}, these cells form a neural ``compass'' maintaining consistent directional representation \citep{Cullen2019}. Intriguingly, the HD system maintains direction representation even without external sensory cues -- a phenomenon known as path integration \citep{McNaughton2006}. This suggests that the HD system functions as a neural integrator updating based on self-motion cues. Theoretical and computational models have proposed that the HD system functions as a continuous attractor network, where the collective activity of HD cells forms a stable ``bump'' of activity that can smoothly move to represent different head directions \citep{Zhang1996, Skaggs1995}. These models often represent the head direction on a ring, reflecting the circular nature of directional space. 

Despite significant progress in understanding the HD system, many questions remain about how its properties emerge from underlying neural circuits and how it interfaces with other components of the brain's spatial navigation system, such as place cells and grid cells \citep{Moser2008,xu2025conformal,zhao2025place}. In this paper, we propose a minimalistic representation model for the HD system that captures its essential features while maintaining computational efficiency and biological plausibility. Our approach is motivated by recent advancements in direction representation learning in high-dimensional spaces \citep{cueva2019emergence,mante2013context,yang2019task,maheswaranathan2019universality}. We leverage the fact that head direction transformations form a representation of the rotation group $U(1)$, acting on a ring of possible head direction representations. We present two versions of the model: a fully connected version and a convolutional version. Both models aim to learn a high-dimensional representation of head direction that exhibits key properties observed in biological HD systems. We demonstrate that our model can learn Gaussian-like tuning profiles for individual cells and produce a representation that exhibits a clear circle geometry when visualized with principal component analysis (PCA). The learned model is capable of accurate path integration. These emergent properties closely match the characteristics of biological HD systems, providing insights into the computational principles that may underlie their function. 

By explicitly incorporating the symmetry of the rotation group and the circular geometry of head direction, our model offers a framework for understanding how these fundamental principles may shape the neural representations in the HD system. This approach not only captures key features of biological HD systems but also provides insights into the role of symmetry and geometry in neural computations for spatial navigation.

\section{Model and Learning}

\subsection{General Framework}

We represent head direction $x \in [0, 2\pi)$ in a continuous $d$-dimensional vector $v(x) \in \mathbb{R}^d$, which is regarded as responses of putative HD cells and subjects to three constraints:

(1) Nonnegativity constraint: $v(x) \geq 0$, reflecting neurons' nonnegative firing rates.

(2)  Unit norm constraint: $|v(x)|^2= \sum_{i=1}^d v_i(x)^2=1$ corresponds to a constant total activity of neurons regardless of direction $x$ (to be one without loss of generality). This implies the direction $x$ is only represented by spatial patterns of neuronal responses rather than summed responses, which has been widely used in neural coding \citep{pouget2003inference,dayan2005theoretical}.

(3) Transformation rule: $v(x+dx) = F(v(x), dx)$, where $F$ is a function describing  changes in the representation $v(x)$ from a change $dx$ in direction. The set of transformations $\{F(\cdot, dx), \forall dx\}$ and the set of representations $\{v(x), \forall x \in [0, 2\pi)\}$ together form a representation of the rotation symmetry group $U(1)$, so that $F(v(x), 0) = v(x)$, and $F(v(x), dx_1+dx_2) = F( F(v(x), dx_1), dx_2)$. Here the addition in $x+dx$ is mod $2\pi$. The transformation rule defines a recurrent neural network $v_t = F(v_{t-1}, dx_t)$ that enables path integration.

% (4) Similarity Structure:
% The inner product between neural response vectors at different directions follows:
% \begin{equation}
% \langle v(x), v(y) \rangle = \exp\left(\frac{\cos(x-y)-1}{\sigma^2}\right).
% \end{equation}
% This constraint establishes that the similarity between neural responses depends only on the angular difference between directions. It creates a von Mises-like distribution of similarity that is consistent with experimental observations. For small angular differences $(x-y)$, this can be approximated as a Gaussian function:
% \(
% \langle v(x), v(y) \rangle \approx \exp\left(-(x-y)^2/2\sigma^2\right).
% \)

\subsection{Model for local motion} 

For local motion $dx$, the first order Taylor expansion gives us 
\begin{equation}
    \begin{aligned}
    v(x+dx) &= F(v(x), dx) \\
        &= F(v(x), 0)  + F'(v(x), 0) dx \\&= v(x) + f(v(x)) dx,
    \end{aligned}\label{eq:motion}
\end{equation}
where $f(v(x)) = F'(v(x), 0)$ is the derivative of $F(v(x), dx)$ with respect to $dx$ evaluated at $dx = 0$. 
This first-order Taylor expansion corresponds to the Lie algebra of the Lie group formed by the transformations $(F(v(x),dx), \forall dx)$. 
For larger motion $dx$, we can also use second-order Taylor expansion. 

To parameterize $f(v(x))$,  here we propose two models, which are the fully connected version and the topographical convolutional version. Our goal is to demonstrate that Gaussian-like head direction tuning profiles emerge regardless of the specific form of $f(v(x))$. Also, we keep both versions of the model as simple as possible, which presents the minimalistic setting. 

\subsubsection{Fully Connected Version}

In the fully connected version, we model local changes in direction as:
\begin{equation}
\label{eq:first-order-linear}
    v(x+dx) = v(x) + B v(x) dx,
\end{equation}
where $B \in \mathbb{R}^{d \times d}$ is a learnable matrix, and $dx \in [-b, b]$ for a small $b > 0$. Here the matrix $B$ enables every dimension of the representation to interact with every other dimension. This flexibility can capture complex, long-range dependencies in the neural code, but it does not impose any specific spatial structure on the arrangement of neurons.

\subsubsection{Topographical Convolutional Version}

The topographical convolutional version of our model explicitly leverages the inherent circular structure of head direction by arranging neurons $v_i$ on a ring. The local update is given by:
\begin{equation}
\label{eq:first-order-conv}
    v(x+dx) = v(x) + B * v(x) dx,
\end{equation}
where $B$ is a learnable convolutional operator, and the symbol $*$ denotes a one-dimensional convolution operation with periodic boundary condition, and $dx \in [-b, b]$ for a small $b > 0$. The convolutional nature of $B$ is expressed as:  
\begin{equation}
    (B * v(x))_i = \sum_{j=-k}^k B_j v_{(i+j) \bmod d}(x),
\end{equation}
where $B_j$ are learnable weights of the convolutional kernel, and $k$ is the kernel size. For each neuron \(i\), the output is computed as a weighted sum of the activities of neurons in its local neighborhood -- specifically, the neurons at positions \(i-k\) through \(i+k\). The modulo operation \(\bmod d\) ensures that the indexing wraps around when \(i+j\) falls outside the range \([0, d-1]\), thereby preserving the circular structure of the representation. This local, circular convolution effectively captures the topological organization of head direction cells, where each neuron's activity is influenced predominantly by its immediate neighbors.

\subsection{Learning Method}

Our model learns two sets of parameters: 

(1) $V$: the representations $v(x)$ for all $x \in \{k\frac{2\pi}{n}, k = 0, ..., n-1\}$, where $n$ is the number of grid points. We denote these $v(x)$ collectively as $V$. For a general continuous $x$, we express $v(x)$ as a linear interpolation between the two nearest grid points. 

(2) $B$: the update matrix or convolution kernel $B$. 

% We define a one-step loss function to train these parameters by minimizing the prediction error of local changes:
We define a loss function to train $B$ and $v$ by minimizing the prediction error of local changes:
\begin{equation}
    \mathcal{L}(V, B) = \mathbb{E}_{x, dx}\left[|v(x+dx)- F(v(x), dx)|^2\right]
\end{equation}
This loss function comes from Equation \ref{eq:motion} and focuses on the accuracy of single-step updates. This one-step loss function can eliminate the need for backpropagation through time, which significantly simplifies the learning process and reduces computational complexity. 

The above loss function can be minimized by projected gradient descent, i.e., after a gradient descent step or a step of Adam optimizer \citep{Kingma2014}, we set all negative elements in each $v(x)$ to 0, and then normalize each $v(x)$ to have norm 1. Expectation $\mathbb{E}_{x, dx}$ can be approximated by uniformly sampling $x$ from $[0, 2\pi)$ and $dx$ from interval $[-b, b]$. 

% In the above loss, we do not enforce similarity structure, and our experiments show that it can already learn main features of head direction system. To enforce similarity structure, we can add an additional loss term 
% \begin{equation}
%      \mathbb{E}_{x, y}\left[\langle v(x), v(y)\rangle - \exp((\cos(x-y)-1)/\sigma^2)\right]^2.
% \end{equation}

\subsection{Model training}
We use Adam optimizer \citep{Kingma2014} to minimize the loss function. The algorithm proceeds as in Algorithm \ref{alg:hd_learning}.
\begin{algorithm}
    \caption{Learning Head Direction Representation}
    \label{alg:hd_learning}
    \begin{algorithmic}[1]
        \Statex \textbf{Input:} Number of directions $n$, dimension $d$, learning rate $\eta$, number of iterations $T$
        \Statex \textbf{Output:} Learned representations $v(x_k)$ and transition function $F$        
        \Statex Initialize $v(x_k)$ for $x_k = k\frac{2\pi}{n}, k = 0, 1, \ldots, n-1$
        \Statex Initialize $B$ (matrix or convolutional kernel)

        \For{$t \gets 1$ to $T$} 
            \State Sample a batch of $(x, dx)$ pairs
            \State Compute the loss $\mathcal{L}$ for the batch
            \State Update $v(x_k)$ and $B$ using gradients of $\mathcal{L}$ and learning rate $\eta$
            \For{$k \gets 0$ to $n-1$}
                \State $v(x_k) \gets \max(v(x_k), 0)$ \Comment{Enforce non-negativity}
                \State $v(x_k) \gets \text{project}(v(x_k))$ \Comment{Project onto unit sphere}
            \EndFor
        \EndFor
        \State \Return $v(x)$ and $B$
    \end{algorithmic}
\end{algorithm}
% \vspace{-5pt}

To achieve a continuous representation, we define $v(x)$ at discrete points $x_k = k\frac{2\pi}{n}$ for $k = 0, 1, \ldots, n-1$, and use linear interpolation for intermediate values:

\begin{equation}
    v(x) = (1-w)v(x_{\lfloor k \rfloor}) + wv(x_{\lceil k \rceil}),
\end{equation}

where $k = \frac{n}{2\pi}x$, $w = k - \lfloor k \rfloor$, and $\lfloor \cdot \rfloor$, $\lceil \cdot \rceil$ denote floor and ceiling functions respectively.

For larger local motion range $b$ (specifically, $b = 20\frac{2\pi}{n}$ in our experiments), we employ a second-order model to capture higher-order dynamics:
\begin{equation}
\label{eq:second-order}
    v(x+dx) = v(x) + B_1 v(x) dx + B_2 v(x) dx^2,
\end{equation}
where $C \in \mathbb{R}^{d \times d}$ is another learnable matrix. This second-order term allows the model to better account for changes over larger directional steps.

\begin{figure}[t!]
    \centering
    \includegraphics[width=0.8\linewidth]{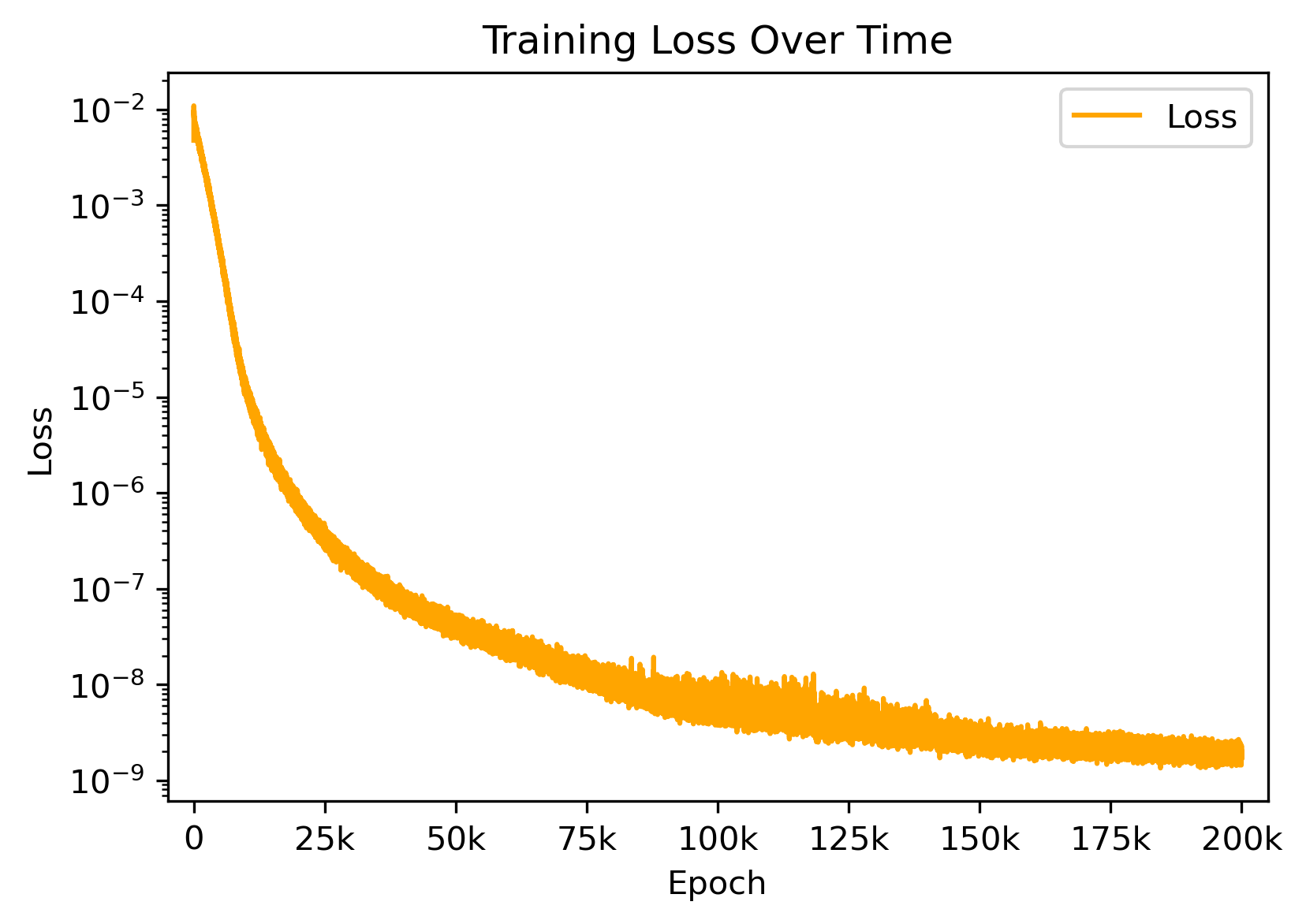}
    \caption{\textbf{Training loss} of an example fully-connected model with dimension $d=20$. We train the model for $200,000$ epochs where it converges.}
    \label{fig:train_loss}
\end{figure}

More specifically in our training process, we use $n=100$ discrete directions. The model was trained for $200,000$ epochs with a batch size of $256$, using an Adam optimizer \citep{Kingma2014} with an initial learning rate of $4e-5$. A learning rate scheduler (ReduceLROnPlateau) is employed with a factor of 0.8 and patience of 5000 epochs to adapt the learning rate during training. For the convolutional model, we use a kernel size of 3. An example training loss curve can be found in \figureref{fig:train_loss}, which shows a stable training process. The model can be trained on a single NVIDIA A6000 GPU in around 5 minutes. 

\section{Experiments and Results}
% report information for one model
% then say we also tried different settings; details in appendix
We conduct a series of experiments to evaluate the performance and properties of our model across various configurations. We explore dimensions $d \in \{10, 20, 50, 100\}$ and local range $b \in \{m\frac{2\pi}{n}, m=2,5,10,20\}$ for both the fully connected and convolutional versions of the model. Here we fix $n=100$ in all experiments.

\begin{figure}[t!]
    \centering
    \includegraphics[width=\linewidth]{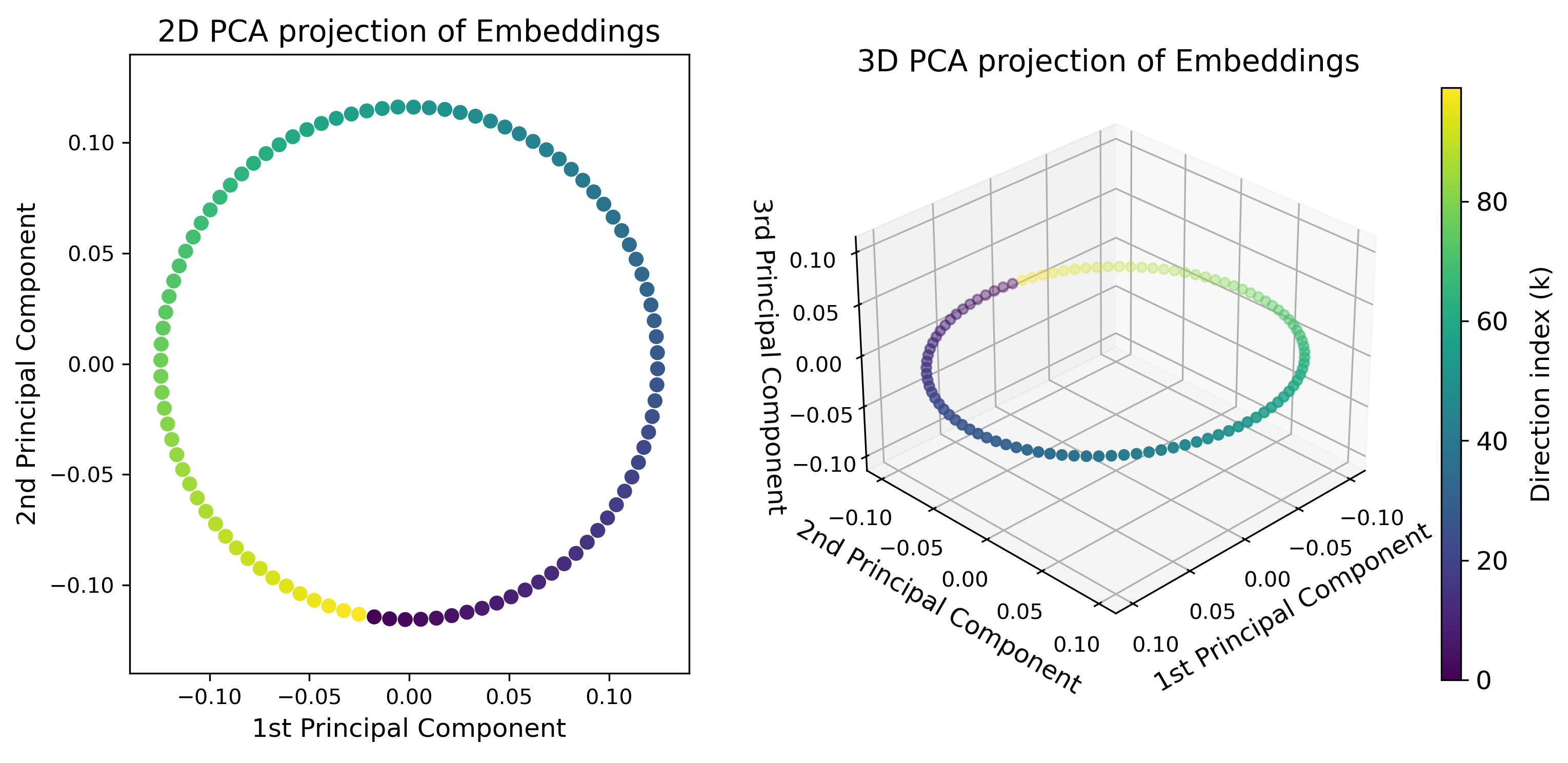}
    \caption{\textbf{2D and 3D PCA visualization} of learned head direction representations. Colors represent the discrete head direction indices from 0 to $n$, corresponding to angles from 0° to 360°. Both PCA plots reveal a clear ring structure, indicating that our model has learned a continuous, circular representation of head direction}
    \label{fig:pca}
\end{figure}

\subsection{Ring Structure in PCA Plot}

We apply Principal Component Analysis (PCA) to the learned representations $v(x)$ across all directions as in \figureref{fig:pca}. The PCA plot of the first two principal components reveal a clear ring structure. This emergent property demonstrates that our model has learned a continuous, circular representation of head direction, mirroring both the topology of the actual direction space and the attractor dynamics observed in biological HD systems. The ring in our PCA plots confirms that even in a high-dimensional firing-rate space, the dominant latent structure reflects a one-dimensional continuum. In other words, small changes in head orientation correspond to smooth movement around the circular manifold, consistent with the view that the HD system operates as a rotationally invariant attractor \citep{Zhang1996}.

Moreover, this ring-like organization is consistent across both model versions, tested dimensions, and local ranges. This result underscores that our high-dimensional representation effectively captures the underlying one-dimensional nature of head direction while providing computational advantages, validating our model's ability to capture essential features of biological head direction systems despite its minimalistic design.

\subsection{Gaussian-like Tuning Profiles}

A defining feature of biological head direction (HD) cells is their unimodal tuning to a specific orientation, resulting in bell-shaped firing curves around a preferred angle \citep{Taube1990a,taube1995head,blair1996visual,McNaughton2006}. In our trained model, we observe a strikingly similar phenomenon: each dimension (or ``cell'') in the learned representation \(v(x)\) shows a maximal response at a particular head direction and exhibits a smooth, approximately Gaussian drop-off for orientations deviating from that preferred angle. 

This emergence of Gaussian-like tuning is critical for a stable and unambiguous directional code. First, it ensures that each cell encodes a unique portion of directional space, distributing orientations across the population \citep{blair1996visual,Taube2007}. Second, Gaussian tuning curves support stable attractor dynamics: small deviations in the system’s activity are naturally pulled back toward the peak direction, mirroring the continuous attractor models that have been proposed to underlie HD cell ensembles \citep{Skaggs1995,Zhang1996}. Third, the unimodal shape of these tuning curves facilitates robust decoding; a small set of maximally responding cells can reliably indicate the current heading, even under noisy or partial input conditions \citep{McNaughton2006}.

\begin{figure}[t!]
    \centering
    \includegraphics[width=0.9\linewidth]{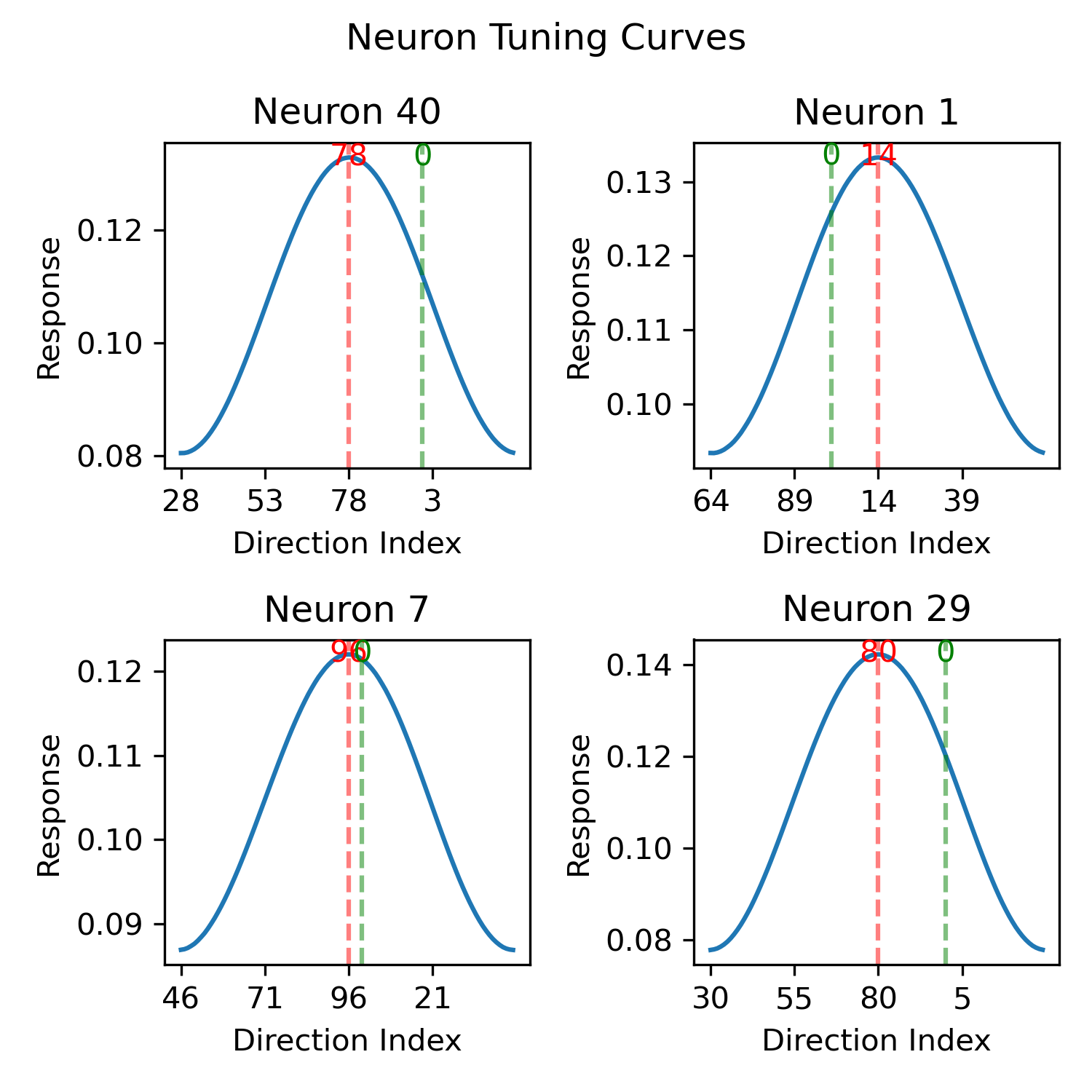}
    \caption{\textbf{Tuning curves of a sample of neurons} from our convolutional model with \(d=20\). This figure, presented as a complement to Figure 4, displays a random selection of neurons to highlight their Gaussian-like tuning profiles. The x-axis spans the full 360° range of head directions, with each curve centered on its neuron’s preferred direction (indicated by the red dotted line) to illustrate the circular nature of the representation. The green dotted line marks the direction index 0.}
    \label{fig:tuning}
\end{figure}

\begin{figure}[t!]
    \centering
    \includegraphics[width=\linewidth]{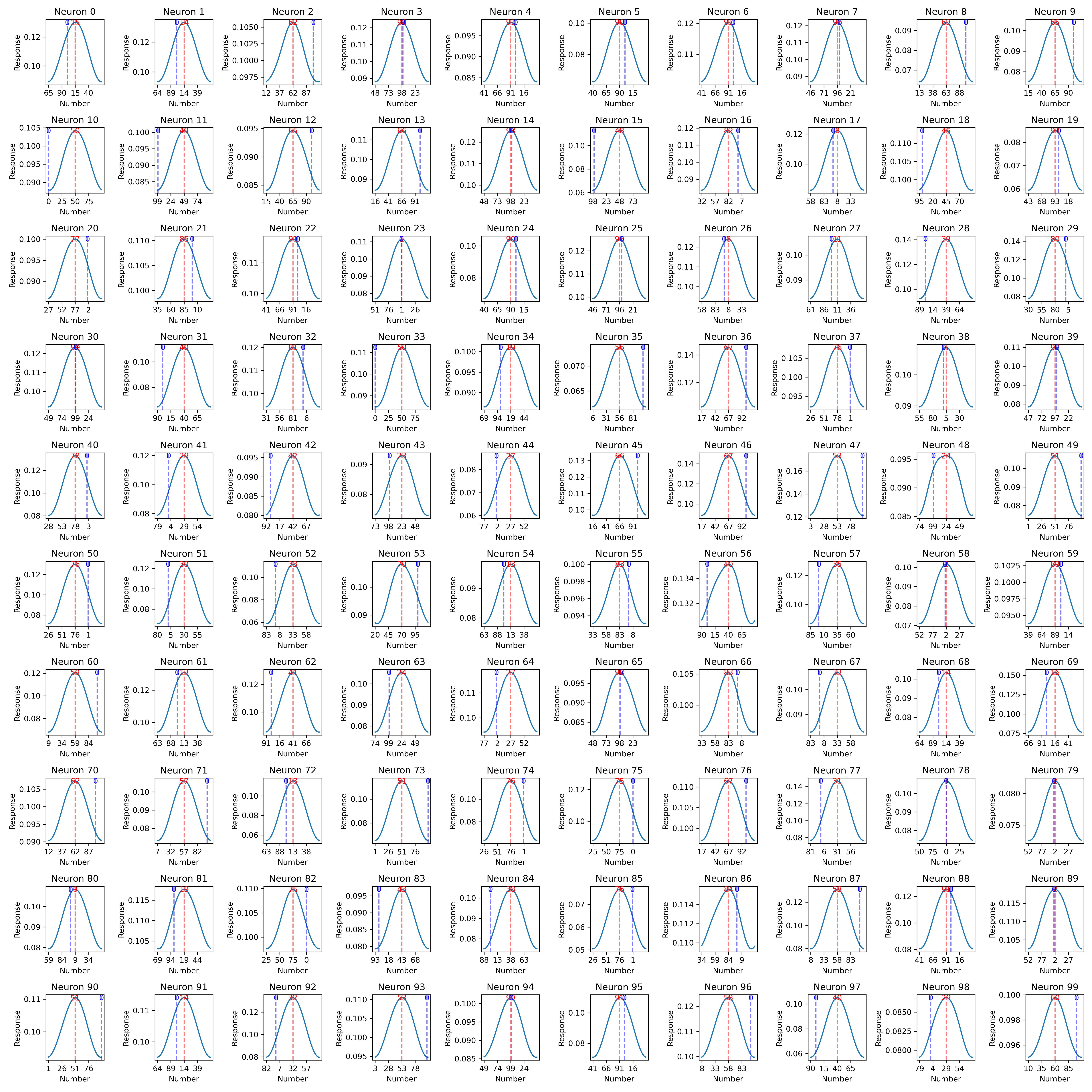}
    \caption{\textbf{Full tuning curves} with $d=100$. We present all $100$ head direction cells, and all tuning curves have Gaussian-like tuning profiles. }
    \label{fig:full_tuning}
\end{figure}

Figure~\ref{fig:tuning} highlights four example neurons from our convolutional model with \(d=20\), each displaying a clear Gaussian-like profile centered on a different preferred direction. To confirm that this behavior holds consistently across the entire network, we further provide the tuning curves of all 100 neurons in our model with $d=100$ in Figure~\ref{fig:full_tuning}. Every dimension in this high-dimensional representation preserves the same unimodal structure, indicating that fundamental symmetry and geometric constraints in the learning process produce stable, biologically realistic firing patterns. These findings closely align with experimental observations of unimodal HD cell tuning in rodents and other mammals.

\subsection{Path integration}

Path integration is a core function of biological head direction (HD) systems, enabling animals to track their heading in the absence of salient external cues by internally summing small changes in orientation \citep{etienne2004path,valerio2012path,McNaughton2006}. In neural terms, it can be seen as sequentially updating an internal representation of direction so that each incremental movement is reflected in the collective firing of HD cells. From a computational standpoint, this corresponds to repeatedly applying a learned update function to a latent state encoding the current heading \citep{Gao2021,Xu2023,xu2023emergence, xu2025conformal}.

To evaluate our model's ability to perform path integration, we begin by assigning an initial head direction $x_0 \in [0, 2\pi)$ and obtaining its latent representation $v_0 = v(x_0)$. We then apply a sequence of directional shifts $(dx_1, dx_2, \ldots, dx_n)$. Given our direction representation function $v(x) \in \mathbb{R}^d$ and the update function $F(v, dx)$, we track the changes in the direction representation by updating the representation for each step $t$ according to 
\begin{equation}
    v_t = F(v_{t-1}, dx_t), t = 1, ..., n
\end{equation}
This procedure ensures that $v_t$ accumulates the integrated effects of all prior increments, thus approximating the final heading after $\sum_{i=1}^t dx_i$. After obtaining the final representation $v_n$, we decode the corresponding direction $x_n$ by maximizing the inner product between $v_n$ and $v(x)$ over all possible direction $x$:
\begin{equation}
    x_n = \arg\max_{x \in [0, 2\pi)} \langle v_n, v(x) \rangle.
\end{equation}

In the decoding phase, we leverage the property that $v_n$ has the highest similarity with $v(x)$ at the true final direction. Such a simple yet effective decoding strategy reflects a population-vector-like scheme often used in neural coding research \citep{georgopoulos1986neuronal}.

To systematically assess performance, path integration is evaluated using two local range scenarios: $b = \frac{2\pi}{n}$ radians, and $b = m\frac{2\pi}{n}$ radians, where $m$ represents the multiple of the basic angular step size used during training. In both cases, the motion of each step $dx_t$ is sampled uniformly from the range $[-b, b]$, thus varying the magnitude of each incremental rotation.  We compare performance both with and without a re-encoding step:
\begin{enumerate}
    \item Without re-encoding: the model simply applies $v_t = F(v_{t-1}, dx_t)$ over multiple steps, causing any representational errors to accumulate. 
    \item With re-encoding: we append an extra decode-then-re-encode procedure at each step. After each step, we first decode $v\rightarrow \hat{x}$ to the 1D head direction angle via $\hat{x} = \arg\max_{x' \in [0, 2\pi)} \langle v, v(x') \rangle$, and then encode $v_t \leftarrow v(\hat{x})$ back to the neuron space intermittently. Since our model is trained in a 1-step manner, this approach aids in rectifying the errors accumulated in the neural space throughout the transformation by resetting the state to a ``clean'' representation.
\end{enumerate}

Table~\ref{table:path} presents the average angular difference (in radians) between the true and estimated directions after multiple steps of path integration. Results are shown for various dimensions $d$ and multiples $m$, and are separated into the two local range conditions. The columns labeled ``local range = $\frac{2\pi}{n}$ error'' report accuracy when each step is sampled from the unit angular step range $\frac{2\pi}{n}$, whereas those labeled ``local range = $m\frac{2\pi}{n}$ error'' show results when each increment is sampled from the range as large as what was used during training. The model maintains low error under both conditions and, importantly, exhibits robust performance at larger step sizes $m\frac{2\pi}{n}$. Notably, despite being trained with a one-step predictive loss function, our model demonstrates remarkable accuracy in multi-step path integration tasks. 

\begin{figure}[t!]
    \centering
    \includegraphics[width=\linewidth]{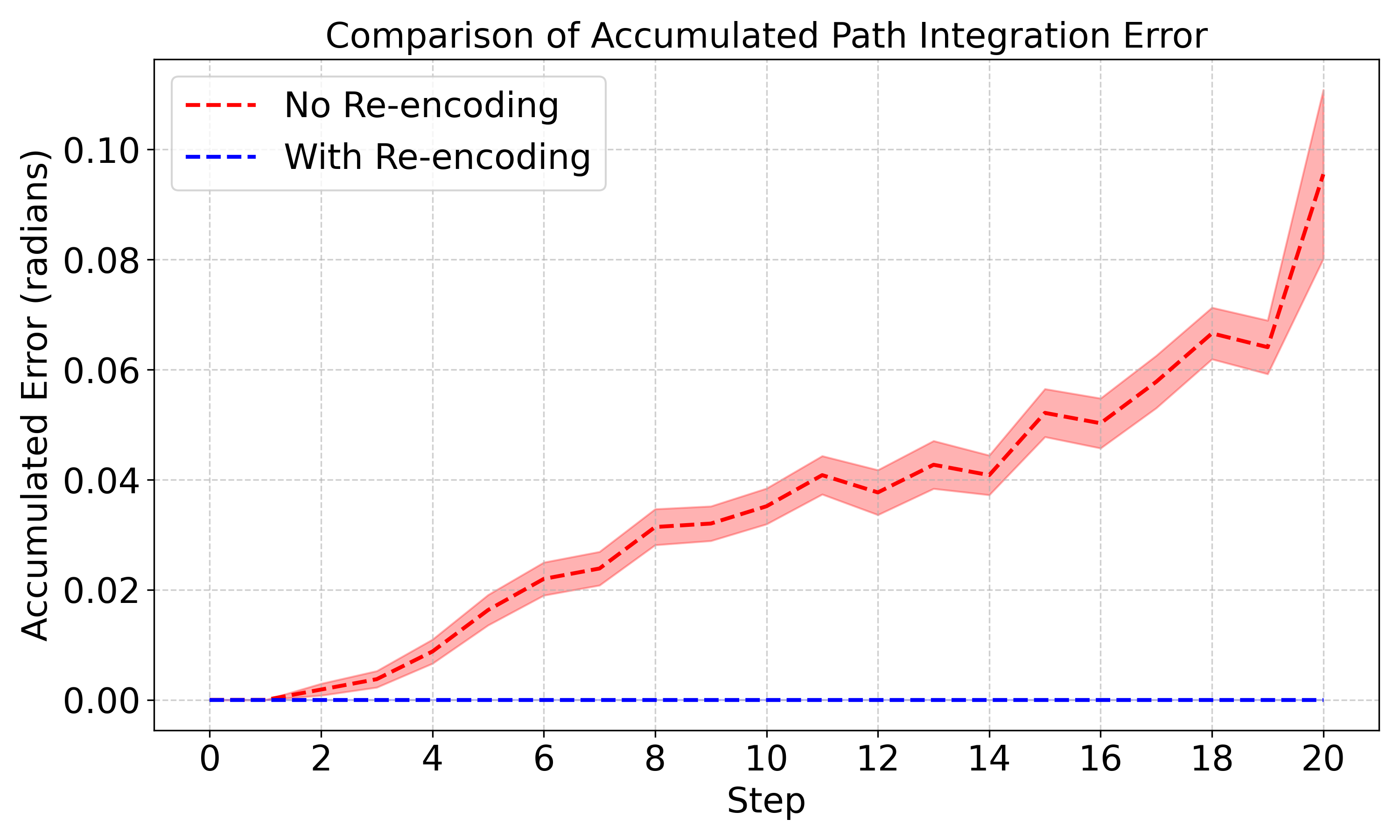}
    \caption{\textbf{Accumulated path integration error} over trajectories of 20 steps with and without re-encoding. We take the model with $d=50$, $m=5$ as an example, i.e., each step $dx$ is sampled from $[- 5\frac{2\pi}{n}, 5\frac{2\pi}{n}]$. Shaded regions indicate standard error across 100 trials. When re-encoding is disabled, we notice an accumulation of path integration error, whereas the re-encoded path remains error-free.}
    \label{fig:cumulative_error}
\end{figure}

\begin{figure}[t!]
    \centering
    \includegraphics[width=\linewidth]{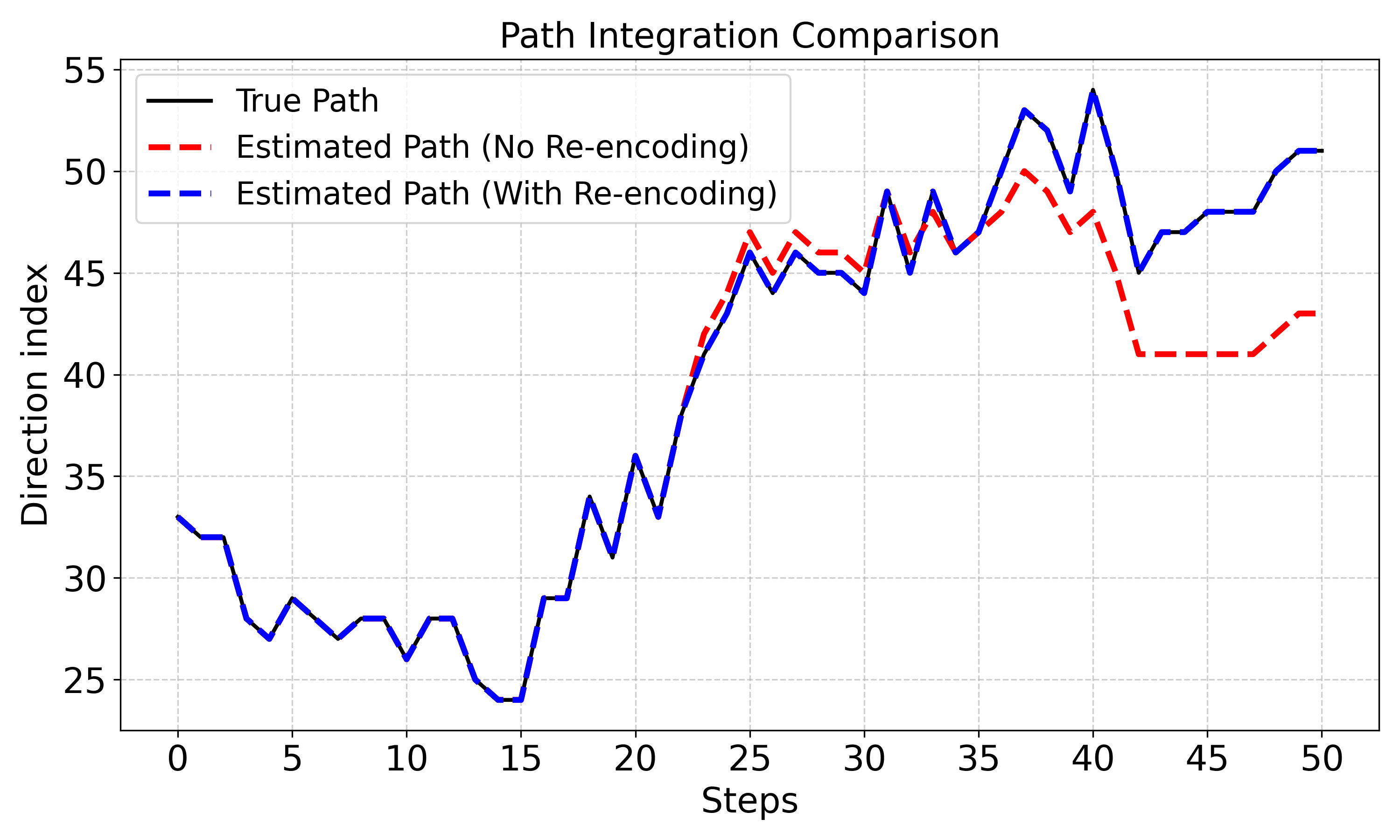}
    \caption{\textbf{Trajectory comparison} over an extended trajectory of 50 steps. We take the model with $d=50$, $m=5$ as an example. The true path (solid black) is overlaid with two estimated paths: one with re-encoding (blue dashed) and one without (red dashed). The re-encoded trajectory aligns perfectly with the true path, while the non-re-encoded estimate gradually deviates.}
    \label{fig:trajectory_plot}
\end{figure}

Figure~\ref{fig:cumulative_error} further shows the effectiveness of re-encoding by comparing the accumulated error over 20 steps in both conditions. The curve without re-encoding shows a gradual increase in error, whereas the error with re-encoding remains low, highlighting the benefit of intermittently snapping the network state back to a ``clean'' representation. Figure~\ref{fig:trajectory_plot} provides an additional view of a 50-step trajectory, in which the re-encoded estimate (blue dashed line) remains essentially indistinguishable from the true path (black solid line), while the non-re-encoded estimate (red dashed line) slowly diverges as cumulative errors accumulate. These results indicate that our minimalistic representation model not only learns smooth, local transformations but also preserves directional accuracy over many updates, generalizing effectively from local updates to global navigation. 

\begin{table}[t]
\centering
\caption{\textbf{Path integration results.} $m$ represents the multiple of the unit angular step size $(\frac{2\pi}{n})$ used for training and evaluation in the larger range scenario. Errors are reported as the average angular difference in radians over all steps in a sequence of length 20. Results are shown for two local motion range scenarios, $b = \frac{2\pi}{n}$ and $b = m\frac{2\pi}{n}$. In both settings, location motions $dx$ are uniformly sampled from the range $[-b, b]$. We train models with $m=2,5,10$ using the first-order transformation as in Equation \ref{eq:first-order-linear} and Equation \ref{eq:first-order-conv} and models with $m=20$ using second-order formulations as in Equation \ref{eq:second-order}.}
\label{table:path}
\resizebox{\linewidth}{!}{%
\begin{tabular}{llr|cccc}
\toprule
{Architecture} & {$d$} & {$m$} & \multicolumn{2}{c}{local range =$\frac{2\pi}{n}$  error (radians)} & \multicolumn{2}{c}{local range = $m\frac{2\pi}{n}$ error (radians)} \\
\cmidrule(lr){4-7}
&  &  & without re-encoding & with re-encoding & without re-encoding & with re-encoding \\
\midrule
\multirow{4}{*}{Fully-connected} & 100 & 2  & 0.000 & 0.000 & 0.069 & 0.000\\
 & 100 & 5  & 0.007 & 0.000 & 0.765 & 0.010\\
   & 100 & 10 & 0.044 & 0.000 & 1.069 & 0.232\\
   & 100 & 20 & 0.356 & 0.000 & 1.362 & 0.179\\
\cmidrule(l){2-7}
   & 50 & 2  & 0.000 & 0.000 & 0.028 & 0.000 \\
   & 50 & 5  & 0.000 & 0.000 & 0.036 & 0.000\\
   & 50 & 10 & 0.008 & 0.000 & 0.872 & 0.167\\
   & 50 & 20 & 0.363 & 0.000 & 0.149 & 0.182\\
\cmidrule(l){2-7}
  & 20 & 2  & 0.000 & 0.000 & 0.008 & 0.000\\
   & 20 & 5  & 0.001 & 0.000 & 0.024 & 0.009\\
   & 20 & 10 & 0.079 & 0.000 & 0.188 & 0.153\\
   & 20 & 20 & 0.385 & 0.000 & 0.186 & 0.154\\
\cmidrule(l){2-7}
 & 10 & 2  & 0.196 & 0.000 & 0.044 & 0.000\\
   & 10 & 5  & 0.000 & 0.000 & 0.047 & 0.014\\
   & 10 & 10 & 0.000 & 0.000 & 1.287 & 0.563\\
   & 10 & 20 & 0.000 & 0.000 & 0.897 & 0.338\\
\midrule
 \multirow{4}{*}{Convolutional} & 100 & 2  & 0.000 & 0.000 & 0.110 & 0.000\\
   & 100 & 5  & 0.002 & 0.000 & 0.832 & 0.000 \\
   & 100 & 10 & 0.056 & 0.000 & 1.252 & 0.184 \\
   & 100 & 20 & 0.035 & 0.000 & 1.322 & 0.107 \\
\cmidrule(l){2-7}
   & 50 & 2  & 0.000 & 0.000 & 0.070 & 0.000\\
   & 50 & 5  & 0.001 & 0.000 & 0.293 & 0.000\\
   & 50 & 10 & 0.008 & 0.000 & 0.866 & 0.167\\
   & 50 & 20 & 0.360 & 0.000 & 1.239 & 0.180\\
\cmidrule(l){2-7}
  & 20 & 2  & 0.000 & 0.000 & 0.088 & 0.000\\
   & 20 & 5  & 0.000 & 0.000 & 0.239 & 0.000\\
   & 20 & 10 & 0.105 & 0.000 & 0.883 & 0.287\\
   & 20 & 20 & 0.035 & 0.000 & 0.144 & 0.180\\
\cmidrule(l){2-7}
   & 10 & 2  & 0.000 & 0.000 & 0.065 & 0.000\\
   & 10 & 5  & 0.002 & 0.000 & 0.125 & 0.014\\
   & 10 & 10 & 0.005 & 0.000 & 0.323 & 0.028\\
   & 10 & 20 & 0.003 & 0.000 & 0.225 & 0.096\\
\bottomrule
\end{tabular}
}
\end{table}

\section{Discussion}

The experimental results demonstrate that our minimalistic model captures key properties of biological HD systems. The emergence of Gaussian-like tuning profiles for individual dimensions of our representation closely resembles the behavior of biological HD cells, suggesting that such tuning properties may arise naturally from fundamental computational principles. The clear ring structure in the PCA plot indicates that our model has learned a continuous attractor representation of head direction, consistent with theoretical models of the HD system.

While head direction is inherently one-dimensional, our model learns a high-dimensional representation, potentially providing computational advantages such as increased robustness to noise and easier integration with other neural systems. Both versions of our model use local update rules that could be implemented by neural circuits, with the convolutional version bearing a particularly close resemblance to the anatomical organization of HD cells. The success of both versions suggests that capturing HD system properties depends more on the overall computational framework than specific architectural details, though the convolutional version may offer advantages in parameter efficiency and biological interpretation.

Our model bridges the gap between detailed biophysical models and more abstract computational approaches. By explicitly incorporating the rotational symmetry of \(U(1)\) and the circular geometry of head direction, our approach offers valuable insights into how these fundamental principles shape neural representations for spatial navigation. In doing so, we aim to contribute to a deeper understanding of the neural basis of spatial navigation and potentially inspire new approaches in both neuroscience and artificial intelligence.

\section{Conclusion}

We present a minimalistic representation model for the head direction system that captures essential features of biological HD systems while maintaining computational efficiency. Our model demonstrates that key properties of HD cells, such as Gaussian-like tuning and a ring structure, can emerge from a simple learning framework based on representing and updating directions in a high-dimensional space.

\section*{Acknowledgment}
Y.W. was partially supported by NSF DMS-2015577, NSF DMS-2415226, and a gift fund from Amazon. 

\bibliographystyle{apacite}
\bibliography{references}

\end{document}